\documentclass[conference]{IEEEtran}
\IEEEoverridecommandlockouts

\usepackage{cite}
\usepackage{amsmath,amssymb,amsfonts}
\usepackage{algorithmic}
\usepackage{graphicx}
\usepackage{textcomp}
\usepackage{xcolor}
\usepackage{booktabs}
\usepackage[utf8]{inputenc}
\usepackage{url}
\usepackage{wasysym} 
\usepackage{balance}

\def\BibTeX{{\rm B\kern-.05em{\sc i\kern-.025em b}\kern-.08em
    T\kern-.1667em\lower.7ex\hbox{E}\kern-.125emX}}

\makeatletter
\newcommand{\linebreakand}{%
  \end{@IEEEauthorhalign}
  \hfill\mbox{}\par
  \mbox{}\hfill\begin{@IEEEauthorhalign}
}
\makeatother

\begin{document}

\title{AI Agent Communication from Internet Architecture Perspective: Challenges and Opportunities}


\author{\IEEEauthorblockN{Chenguang Du}
\IEEEauthorblockA{\textit{Zhongguancun Laboratory}\\
Beijing, China \\
ducg@zgclab.edu.cn}
\and
\IEEEauthorblockN{Chuyi Wang}
\IEEEauthorblockA{\textit{Tsinghua University}\\
Beijing, China \\
wangchuy21@mails.tsinghua.edu.cn}
\and
\IEEEauthorblockN{Yihan Chao}
\IEEEauthorblockA{\textit{Zhongguancun Laboratory}\\
Beijing, China \\
chaoyh@zgclab.edu.cn}
\linebreakand
\IEEEauthorblockN{Xiaohui Xie}
\IEEEauthorblockA{\textit{Tsinghua University}\\
Beijing, China \\
xiexiaohui@tsinghua.edu.cn}
\and
\IEEEauthorblockN{Yong Cui}
\IEEEauthorblockA{\textit{Tsinghua University}\\
Beijing, China \\
cuiyong@tsinghua.edu.cn}
}

\maketitle

\begin{abstract}
The rapid development of AI agents leads to a surge in communication demands. Alongside this rise, a variety of frameworks and protocols emerge. While these efforts demonstrate the vitality of the field, they also highlight increasing fragmentation, with redundant innovation and siloed designs hindering cross-domain interoperability. These challenges underscore the need for a systematic perspective to guide the development of scalable, secure, and sustainable AI agent ecosystems.
To address this need, this paper provides the first systematic analysis of AI agent communication from the standpoint of Internet architecture—the most successful global-scale distributed system in history. 
Specifically, we distill decades of Internet evolution into five key elements that are directly relevant to agent communication: scalability, security, real-time performance, high performance, and manageability. We then use these elements to examine both the opportunities and the bottlenecks in developing robust multi-agent ecosystems.
Overall, this paper bridges Internet architecture and AI agent communication for the first time, providing a new lens for guiding the sustainable growth of AI agent communication ecosystems.
\end{abstract}


\section{Introduction}

An \textit{AI agent} is an entity capable of autonomous perception, decision-making, and action. Supported by foundational computing, storage, and communication infrastructures, agents are proliferating across diverse forms, ranging from autonomous applications to physical robots. With the advent of large language models (LLMs), the capabilities of AI agents have been significantly enhanced, greatly expanding the scope of multi-agent collaboration.

As this trend accelerates, a variety of communication technologies and protocols have rapidly emerged. For example, Google proposed the Agent2Agent Protocol (A2A)~\cite{a2a_google}, while open-source contributors developed the Agent Network Protocol (ANP)~\cite{anp}, both aiming to enable cross-agent interoperability and coordination. This vibrant wave of innovation underscores the vitality of the field but also reveals growing fragmentation. Two issues stand out in particular. First, innovation redundancy: leading companies and talented developers are frequently duplicating efforts, resulting in wasted resources. Second, value fragmentation: while the core purpose of agent communication is to enable cross-domain interoperability and collaboration, today’s siloed designs often hinder seamless interaction across systems. These challenges highlight the urgent need for systematic perspectives and established design principles. A unified analytical lens is essential to coordinate diverse efforts, reduce fragmentation, and promote the healthy and sustainable development of the agent ecosystem.

Building on this perspective, this paper provides the first systematic analysis of AI agent communication from the standpoint of Internet architecture. Instead of focusing on isolated protocols or platforms, we draw on decades of Internet evolution to extract key elements for understanding emerging agent ecosystems. Using this approach, we make three main contributions: (1) we identify five key elements—scalability, security, real-time performance, high performance, and manageability—that capture the fundamental challenges of agent communication; (2) we examine these elements to highlight both the bottlenecks and the opportunities in developing robust multi-agent ecosystems; and (3) we extend the discussion to standardization, identifying future opportunities for agent communication standards.

The remainder of this paper is organized as follows. Section II reviews the evolution of agent communication technologies. Section III analyzes the key challenges and opportunities of agent communication through the lens of Internet architecture. Section IV discusses potential standardization pathways at both the transport and application layers, and examines the strategic competition for agent communication entry points. Finally, Section V concludes the paper.

\section{Existing Agent Communication Technologies}


\begin{table*}[htbp]
\centering
\caption{Comparison of Agent Communication Protocols/Frameworks across six attibutes.\\ \scriptsize{\Circle : not supported, \LEFTcircle : partially supported, \CIRCLE : well supported.}}
\resizebox{.96\textwidth}{!}
{\begin{tabular}{|l|c|c|c|c|c|c|l|l|}
\hline
\textbf{Framework / Protocol} & \textbf{Disc.} & \textbf{Auth.} & \textbf{Cap. Desc.} & \textbf{Neg.} & \textbf{Exec.} & \textbf{Audit.} & \textbf{Strengths} & \textbf{Weaknesses} \\
\hline
KQML / FIPA-ACL & \LEFTcircle & \LEFTcircle & \CIRCLE & \CIRCLE & \LEFTcircle & \Circle 
& Rich speech-act semantics, formal negotiation & No Internet-scale identity, no governance, weak runtime \\
OAuth / OIDC    & \LEFTcircle & \CIRCLE     & \LEFTcircle & \Circle & \Circle & \LEFTcircle 
& Standardized delegated auth, identity claims & Not agent-native, no execution, limited discovery \\
DID  & \CIRCLE     & \CIRCLE     & \Circle & \LEFTcircle & \Circle & \LEFTcircle 
& Self-sovereign, verifiable identity and discovery endpoints & Limited semantics, runtime and governance challenges \\
AutoGen  & \LEFTcircle & \LEFTcircle & \LEFTcircle & \Circle & \CIRCLE & \LEFTcircle 
& Orchestration via role-based dialogue & Platform-bound, weak discovery/auth, limited audit \\
Dify    & \LEFTcircle & \LEFTcircle & \LEFTcircle & \Circle & \CIRCLE & \LEFTcircle 
& Graphical orchestration, observability & Proprietary SDKs, inconsistent semantics, poor cross-platform \\
MCP   & \LEFTcircle & \LEFTcircle & \CIRCLE & \LEFTcircle & \CIRCLE & \CIRCLE 
& Unified model-tool interface, context injection & Externalized security, needs shared audit/governance \\
A2A / ANS & \CIRCLE  & \LEFTcircle & \LEFTcircle & \LEFTcircle & \LEFTcircle & \CIRCLE 
& AgentCard metadata, ANS naming & Immature ecosystem, modest semantics, scaling issues \\
ANP & \CIRCLE   & \LEFTcircle & \CIRCLE & \CIRCLE & \LEFTcircle & \CIRCLE 
& Integrated discovery, description, negotiation, session security & Deployment immaturity, governance overhead \\
Agents SDK  & \LEFTcircle & \LEFTcircle & \CIRCLE & \LEFTcircle & \CIRCLE & \LEFTcircle 
& Maps protocol to runtime, tool registration, guardrails & Platform-specific, weak interoperability, limited audit \\
\hline

\end{tabular}}
\end{table*}

This section presents a structured and historically grounded review of agent-to-agent communication technologies. The organizing principle is the progressive enhancement of interoperability capabilities: from early semantic formalisms, to web identity and engineered bearers, and eventually to protocolized interconnection in the large-model era. Rather than following a linear chronology, the analysis is problem-driven, reconstructing successive technical attempts to address discovery, authentication, capability description, negotiation, execution, and auditability across various domains. The review is divided into two major periods: the pre-LLM era, where technical primitives laid conceptual and infrastructural groundwork yet left interoperability gaps unresolved, and the era of LLMs, where orchestration frameworks and emerging protocols have reframed the boundaries and bottlenecks of cross-ecosystem interaction. Each subsection situates technologies within their problem context, tracing both achievements and enduring limitations, thereby generating evidence that informs the standardization pathways discussed in later sections.

\subsection{Pre-LLM Era: Foundation of Communication Framework}


This subsection revisits the foundational primitives established prior to large models, providing an analysis of their capacities and limitations. Speech-act semantics translated intent into executable capability descriptions, providing expressiveness; the Web identity and semantic stack bound these descriptions to resolvable entities, ensuring reference and traceability; engineered carrier models encapsulated them as messages, tokens, or API calls, securing reliable delivery; and decentralized identifiers reduced dependence on single trust anchors, enhancing cross-domain verifiability. Together, these form a causal chain: expression, binding, transmission, verification, but unresolved challenges in Internet-scale addressing, cross-domain trust, and governance show that no single primitive can sustain interoperability. The achievement of Internet-scale coordination requires the architectural composition of multiple primitives.

\subsubsection{Semantic Communication Foundations}

KQML~\cite{finin1994kqml} and FIPA-ACL~\cite{fipa2002fipa} introduced the ``speech act'' paradigm and session management into engineering practice, establishing the semantic framework for multi-agent interaction. These protocols treated messages as intentional behavioral units (performatives) that carry performative verbs such as ``request'', ``inform'', and ``query'', while defining conversation turns, commitments, and semantic constraints in dialogue protocols. This paradigm elevated complex collaboration from program interfaces to semantic contracts, providing operational primitives for capability description, negotiation, and verification. However, these protocols failed to address Internet-level addressing, verifiable identity, and governance mechanisms, making them insufficient for large-scale cross-domain interconnection and remaining largely in research prototypes.

\subsubsection{Web/Identity and Semantic Infrastructure}

OAuth 2.0/OIDC~\cite{hardt2012rfc} and JSON-LD~\cite{kellogg2019json} solidified authorization/identity and Web-native semantic expression as de facto standards, providing a reusable foundation for agents to achieve ``verifiable identity'' and ``linkable semantics'' in cross-domain environments. OAuth 2.0 provided operational processes for delegated authorization, while OpenID Connect added identity claims and basic trust interactions. JSON-LD treated semantic context as a first-class element of data, enabling machine-parsable semantic annotations during cross-domain data flow. These technologies were not designed specifically for agents, but provided direct toolchains for agent capability description, credential exchange, and semantic alignment.

\subsubsection{Engineering Implementation: RPC and Event Models}

The ``RPC extremes'' of gRPC~\cite{wang1993grpc} and JSON-RPC~\cite{samsel2013web}, along with CloudEvents' unified event meta-model, engineered the bearer layer of multi-agent communication into deliverable runtime foundations. Strongly typed gRPC suited high-throughput, low-latency service mesh scenarios, while lightweight JSON-RPC attracted developers with lower integration costs in heterogeneous environments. CloudEvents standardized the event stream metadata to enable consistent event envelopes for publishers and consumers. For multi-agent systems, engineered bearers provided reliable communication, observability, and performance guarantees, incorporating task scheduling, tool invocation, and log tracking into measurable runtime environments.

\subsubsection{Verifiable Identity Enhancement}

The decentralized identifier (DID)~\cite{reed2020decentralized} combines the identifier, public key, verification method, and service endpoint into independently resolvable ``DID documents''. This provides a practical path to establish trusted identity and discovery endpoints in multi-autonomous domain environments that lack a shared root of trust. DID shifts identity from platform accounts to self-sovereign units that support cross-domain verification. It also enables endpoint discovery bound to identity claims, laying the foundation for verifiable discovery despite governance and performance challenges in large-scale deployment.

\subsection{The Era of LLMs: From Platformization to Protocolized Interoperability }
This subsection analyzes platformization and nascent protocolization in the large language model era, encompassing orchestration schemes, MCP-style~\cite{anthropic_mcp} abstraction, AgentCard~\cite{a2a_google} /ANS~\cite{huang2025agent} proposals, and ANP~\cite{anp} meta-protocols. These efforts extend earlier primitives while addressing new constraints: platform silos impede interoperability, auditability safeguards traceability, and credential-recognition mechanisms sustain trust in multi-actor collaboration. In doing so, they amplify expressiveness, linkability, and delivery guarantees, and advance interface-level standardization toward Internet-scale agent cooperation.


\subsubsection{Platformization of Orchestration}
LLM-centric multi-agent~\cite{bellifemine2000developing} frameworks made ``conversation-as-orchestration'' mainstream, maturing orchestration patterns including supervisor-executor, debate-style reasoning, and tool chain invocation, significantly lowering thresholds for building collaborative agent systems. Frameworks like AutoGen~\cite{wu2024autogen} decomposed complex tasks into role-based dialogue participants, allowing decision-making, search, and tool invocation through message interaction. However, most frameworks relied on platform-provided storage, logging, and routing mechanisms, with discovery, authentication, and credentials often being platform-specific or manually configured, limiting cross-organization interoperability. This gap underscores the importance of developing standardized identity, communication, and coordination protocols, an area increasingly emphasized in recent systemization-of-knowledge studies.


Dify~\cite{dify}, LangGraph~\cite{langgraph}, and others introduced graphical orchestration and finite-state machines into production environments, significantly improving controllability and observability while exposing ``platform silo'' and interface privatization barriers. Graphical orchestration explicitly represented complex tasks as node-edge structures where nodes represent agents/tools and edges represent data/control flow. Through visualization and finite-state machine constraints, platforms improved fault recovery and debugging efficiency. However, proprietary SDKs and metadata models in the interface layers led to inconsistent capability descriptions and context models between platforms, increasing cross-platform integration costs. This fragmentation highlights the need for shared orchestration ontologies and interoperability standards to enable heterogeneous agent ecosystems to collaborate more effectively.


\subsubsection{Protocolization of Interfaces}

Model Context Protocol (MCP)~\cite{anthropic_mcp} standardized access between models and external tools/data sources, reducing platform adaptation costs and representing milestones in elevating SDK layer capabilities to cross-ecosystem protocols. MCP abstracted model-tool interactions into unified capability descriptions, handshake processes, and context injection contracts, enabling dynamic requests and secure use of external tools at runtime. As the ``USB-C for AI'' analogy suggests, MCP's value lies in reducing engineering friction in tool access between different platforms or models, though success depends on common authentication, quota, and audit specifications. Recent studies also emphasize that without converging governance mechanisms and standardized auditing practices, such protocols risk reproducing existing fragmentation across ecosystems.


A2A's proposed AgentCard (e.g., /.well-known/agent.json) and ANS (Agent Name Service)~\cite{a2a_google, huang2025agent} aim to expose ``who can do what'' and ``how to contact it'' in Web-native and verifiable forms, bringing discovery/naming problems into governable scope. AgentCard provides self-descriptive metadata (provider, endpoint, capability list, authentication/signature information), enabling discovery services to achieve trusted indexing without centralized directories. ANS introduces DNS-like naming and resolution mechanisms that support capability-oriented search and aggregation. Together, they transform traditional ``finding people'' problems into engineering problems of finding autonomous capability units. Previous work in service discovery and naming has highlighted similar tensions between decentralization, scalability, and trust management. More recent studies in large-scale distributed systems also show that integrating semantic descriptions with resolution infrastructures is key to enabling interoperability across heterogeneous agent environments.


Agent Network Protocol (ANP)~\cite{anp} merges Discovery and Description (represented by JSON-LD, schema.org) with a Communication Meta-Protocol, incorporating name, capability, credential, and dialogue
negotiation into unified composable models for cross-ecosystem end-to-end interoperability. ANP's core idea is explaining not only ``what can be done'' but also ``how to agree on interaction input-output semantics, authorization boundaries, and failure compensation strategies.'' Therefore, ANP defines capability description vocabularies, negotiation message formats, session security and encryption strategies, as well as failure semantics and fallback schemes. If ANP forms mutually recognized contracts across platforms, it will significantly reduce cross-domain integration costs.


\subsubsection{Execution and Governance Runtime}

Agents SDK~\cite{openai_agents} maps protocol-level descriptions to executable toolsets and runtime primitives (tool registration, handoff, guardrails, state management), implementing ``protocolized discovery and semantic negotiation'' as orchestrable and observable application processes. However, cross-protocol mutual recognition and auditing remain shortcomings. In production settings, the SDK converts high-level negotiations into concrete invocation flows. Reliability is ensured through error handling, compensation, and monitoring. Unified access to external tools, token management, and request routing is also provided. However, SDK implementations typically favor specific platforms, making mutual recognition across platforms difficult, along with challenges in maintaining auditable traceability chains and protecting sensitive contexts at runtime.




\subsection{Summary and Insights}

The historical trajectory of agent communication reveals a recurring tension between expressive semantics and practical interoperability. Early frameworks such as KQML and FIPA were characterized by strong formal expressiveness, but their adoption was hindered by deployment barriers. Web-native standards such as OAuth and JSON-LD were introduced to improve trust and portability, yet fragmentation across domains was left unresolved. Transport-oriented efforts such as gRPC and CloudEvents were designed to enhance delivery efficiency, but shared semantics and governance were largely neglected. Emerging proposals, including MCP, AgentCard/ANS, and ANP, aim to restore large-scale interoperability but are still in formative stages. Taken together, these developments show that incremental advances have addressed isolated dimensions of the interoperability problem, yet no architecture has reconciled semantics, trust, and runtime integration. Fragmented frameworks also create reliability gaps, including message loss, inconsistent state propagation, and failures in coordinated behavior, limiting sustainable ecosystem growth. Building on these observations, the next section develops an analytical framework to systematically evaluate the fundamental challenges and potential development paths of agent communication.

\section{Challenges and Opportunities in AI Agent Communication}

As AI agents scale across heterogeneous and cross-domain environments, fragmented frameworks reveal significant gaps in interoperability and reliability, limiting sustainable ecosystem growth. 
To this end, we draw on the principles of Internet architecture to extract key elements.
These elements capture the fundamental challenges and point toward potential development paths for emerging agent communication development.

\subsection{Internet-Inspired Key Elements}

To systematically analyze the agent ecosystem, we adopt the architecture of the Internet not merely as a historical analogy, but as a source of time-tested design elements. The Internet’s success in connecting a global, heterogeneous network was contingent on mastering a set of fundamental design challenges. Its solutions, refined over decades of operation and evolution, provide a time-tested lens through which to evaluate the nascent field of agent communication.

Specifically, we distill the Internet’s architectural experience into five key elements that closely parallel the challenges agents face today: scalability, to manage explosive growth in entities and interactions; security, to establish trust among autonomous and often unfamiliar entities; real-time and high performance, to enable complex, time-sensitive collaboration; and manageability, to ensure operational stability and observability in a dynamic environment. These five pillars represent the foundational issues that will determine the future viability of an interconnected ``Internet of Agents". By examining the emerging landscape of agent communication through these dimensions, we can move beyond ad-hoc observations to systematically identify foundational problems and uncover architecturally-sound opportunities for future development.

\subsection{Scalability}

\subsubsection{Definition of Scalability and Internet Implementation}

Scalability refers to the ability of a system, network, or process to operate efficiently and meet performance requirements as its scale expands. In complex communication systems, when nodes and data surge and interaction relationships become increasingly complex, scalability is crucial: it ensures that the system can respond to load growth without crashing.

The Internet architecture has attached great importance to scalability issues since its birth. Through carefully designed mechanisms to support large-scale expansion, it has successfully evolved from the initial small experimental network connecting several computers to today's vast network connecting billions of devices worldwide. Among them, the design ideas of packet switching, regional autonomy, and layered architecture have played important roles.

\subsubsection{Scalability Challenges}

With the rise of multi-agent systems, their communication architecture faces scalability challenges similar to but more severe than those of the early Internet. These challenges can be summarized into two main categories:

\textbf{Communication Explosion}.  
Multi-agent collaboration inherently requires frequent information exchange. As the number of agents and tool-based agents continues to grow rapidly, and as their states and capabilities change dynamically, communication overhead can quickly spiral out of control. Without effective organization, systems may easily degenerate into full-mesh patterns, where each agent interacts with all others, leading to quadratic $O(N^2)$ growth in communication links. This uncontrolled communication explosion undermines efficiency and threatens overall system stability.

\textbf{Discovery Dilemma}.  
Another key bottleneck lies in agent discovery, which departs fundamentally from the Internet’s DNS. Instead of exact name-to-address resolution, discovery is similar to a search engine, matching user intent to ranked capabilities through semantic resolution. 
This raises several intertwined issues. First, the challenge of matching demand and capability is no longer a simple lookup problem; it requires contextual understanding, assessment of quality, and interpretation of user intent to rank multiple dynamic capabilities. Second, discovery must also incorporate mechanisms for conflict resolution and misuse prevention. For instance, in weather forecasting, different providers’ agents may simultaneously return contradictory outputs, making trust management and conflict resolution indispensable. Third, the assignment and management of agent identifiers become critical as the ecosystem scales. Agents require unique and verifiable IDs, but whether these should be allocated by centralized platforms or dynamically generated based on content and capabilities remains an open question, with significant implications for both scalability and governance.

\subsubsection{Scalability Development Opportunities}

Drawing lessons from Internet design, future agent communication systems may embrace the following development opportunities:

\textbf{Global Agent Identity}.  
A Global Agent Identity System is required to issue and manage verifiable, resolvable, and unique IDs for all agents. Unlike traditional addressing, these identifiers would be designed to embed descriptive metadata about an agent's capabilities and roles. A hierarchical parsing and addressing scheme within GAIS would then enable both reliable identification and intent-driven routing in large-scale, heterogeneous environments.

\textbf{Capability Discovery}.
A Capability Discovery System would establish a layered, distributed service to replace centralized, omniscient registries. Functioning as a DNS for the agent era, this system would resolve queries for ``what can be done" (finding capabilities) rather than ``where is it" (finding addresses). Through mechanisms like caching and hierarchical aggregation, a CDS would enable agents to dynamically locate appropriate peers based on their intentions and functional descriptions, rather than relying on static identifiers.

\textbf{Autonomous Agent}.
Inspired by Internet Autonomous Systems, agents can be grouped into cooperative clusters that handle internal communication efficiently while exposing standardized boundary interfaces for cross-cluster interaction. This transforms the uncontrolled explosion of connections into a manageable engineering problem of intra-group optimization and inter-group standardization.

\subsection{Security}

\subsubsection{Definition of Security and Internet Implementation}

Security refers to a communication system’s ability to protect against malicious attacks, unauthorized access, and misuse, while ensuring confidentiality, integrity, and availability of data and services. 

The Internet has accumulated a layered security architecture over decades, including authentication and authorization frameworks (e.g., OAuth 2.0), secure transmission protocols (e.g., TLS/SSL), and auditability mechanisms such as digital signatures and Public Key Infrastructure (PKI). These mechanisms enable reliable trust management across billions of heterogeneous entities, but they also expose important limitations when extended to the more dynamic and semantically complex world of agent communication.

\subsubsection{Security Challenges}

With the proliferation of autonomous agents, security challenges expand beyond conventional network threats, encompassing identity, delegation, data confidentiality, and systemic governance. These challenges can be grouped into three categories:

\textbf{Authentication and Authorization}.  
Multi-agent systems demand not only identity authentication but also capability authentication. An agent may represent a user, an organization, or even another agent, creating multi-layered trust chains. This raises three intertwined issues.
First, identity authentication is essential to prevent impersonation attacks in which malicious agents attempt to disguise themselves as trusted service providers. Second, capability authentication must ensure that an agent truly possesses the skills or permissions it claims, rather than merely advertising false capabilities. Third, delegated authorization becomes critical in complex workflows, where permissions are passed from a user to an agent, and potentially onward to other agents. Such delegation increases the risk of privilege escalation or misuse, making robust authorization mechanisms indispensable for secure collaboration in large-scale agent ecosystems.


\textbf{Communication and Data Transmission}.  
The dynamic and high-frequency communication of agents introduces unique risks in data transmission.
One critical threat is the man-in-the-middle (MITM) attack: without strong end-to-end encryption, adversaries may intercept, manipulate, or replay the messages exchanged among agents. Another challenge concerns context leakage. Since agents rely on sharing contextual information to collaborate, excessive disclosure can expose sensitive data, such as personal schedules or proprietary organizational knowledge. Addressing these risks requires secure communication protocols that balance information sharing with confidentiality and integrity guarantees.


\textbf{Governance and Accountability}.  
Open multi-agent ecosystems pose significant challenges in governance and traceability, which can be broadly categorized into three aspects. First, abuse prevention is critical, as agents may be exploited for misinformation, spam, or automated attacks. A particularly concerning risk is discovery abuse, where agents manipulate ranking algorithms in ways analogous to adversarial practices in Search Engine Optimization (SEO). By falsifying capabilities, padding descriptions with irrelevant metadata, or fabricating usage statistics, agents can gain disproportionate visibility. Such ``agent-oriented SEO manipulation'' threatens the integrity of discovery services and erodes ecosystem trust. Second, content and behavior management becomes essential, since autonomous agents can generate harmful or misleading outputs that require effective moderation strategies. Third, traceability and non-repudiation remain pressing concerns: when agents act erroneously or maliciously, assigning responsibility is difficult without tamper-proof audit trails and accountability mechanisms.


\subsubsection{Security Development Opportunities}

Drawing on Internet security experience while addressing agent-specific complexities, several development opportunities can be identified:

\textbf{Agent Identity Authentication}.  
Establishing a standardized identity layer with globally unique identifiers for agents is critical. These IDs should be verifiable and resolvable. Embedding metadata that describe an agent’s skills and capabilities into credential certificates enables validation of both identity and competence.

\textbf{Authorization Framework}.  
Traditional user–service authorization must evolve into multi-entity frameworks that account for both user-to-agent and agent-to-agent interactions. In the first case, user-to-agent authorization requires clear mechanisms for scoping permissions and enabling timely revocation, ensuring that users retain effective control over what their agents can do. In the second case, agent-to-agent delegation necessitates cryptographically enforced constraints on the scope of delegated rights, so as to prevent privilege escalation when permissions are passed along through chained grants. Together, these mechanisms form the foundation for secure and accountable authorization in large-scale multi-agent ecosystems.

\textbf{Minimal Context Sharing}.  
To mitigate information leakage, communication protocols should adopt the “least context exposure” principle, with fine-grained data isolation, content redaction, and privacy-preserving techniques ensuring agents exchange only what's strictly necessary.

\textbf{Trusted Governance and Accountability Mechanisms}.  
Ensuring integrity in agent ecosystems requires mechanisms that combine trustworthy discovery with accountable governance. Discovery should move beyond unverifiable claims by embedding Verifiable Capability Credentials into the process, allowing skills and permissions to be cryptographically proven. To counter manipulation such as ``Agent SEO”, reputation systems can rank agents based on historical performance, feedback, and auditable logs—functioning as a ``PageRank for Agents”. In parallel, governance can be reinforced through digital signatures and distributed ledger technologies, ensuring tamper-resistant records, auditability, and non-repudiation. Together, these mechanisms foster a meritocratic and accountable discovery environment where trust, not manipulation, determines visibility.

\subsection{Real-time}
\subsubsection{Real-time Definition and Internet Implementation}
Real-time capability refers to a system's ability to respond within strict time constraints when receiving inputs or events. The core metric is low end-to-end latency and strong state consistency. In practical scenarios such as live streaming, online gaming, or AR/VR applications, real-time performance directly affects user-perceived responsiveness (e.g., video stuttering or control lag). For autonomous vehicles, real-time communication between sensors and control units must occur within milliseconds to ensure safe navigation.

The Internet has achieved real-time capabilities through both protocol-layer and architecture-layer optimizations. At the protocol level, UDP provides lightweight, connectionless transmission suitable for latency-sensitive applications (e.g., VoIP, live streaming). QUIC, built on UDP, integrates encryption, multiplexing, and 0-RTT connections, significantly reducing handshake latency (e.g., YouTube Live). WebRTC enables peer-to-peer low-latency communication in browsers, supporting multiparty conferencing via Pub/Sub mechanisms. At the architectural level, Content Delivery Networks (CDN) cache static and dynamic content at global edge nodes, reducing access latency (e.g., Netflix video loading time improved by 50\%). Edge computing further brings computation closer to end users (e.g., MEC for IoT or AR rendering), avoiding round trips to central clouds. In addition, intelligent routing techniques such as SDN dynamically optimize paths based on bandwidth, delay, and packet loss metrics.

\subsubsection{Real-time Challenges}
In multi-agent systems, real-time demands are particularly stringent.

\textbf{State synchronization latency}. In highly dynamic environments, agents must synchronize critical state information at millisecond-level precision. Delays may cause inconsistent worldviews among agents. For example, warehouse robots competing for the same shelf due to unsynchronized positions may collide, and distributed UAV swarms may lose formation if their locations are not consistently updated. In autonomous driving scenarios, vehicles must share position and velocity data within 10ms to prevent collisions—a requirement TCP cannot meet due to its retransmission overhead.

\textbf{Protocol design dilemma}. Agent communication must simultaneously support ultra-low latency (e.g., control signals within $<$10ms) and structured data processing (e.g., semantic-rich messages containing priorities, logic, or environmental metadata). For instance, a video-analysis agent must stream frames in real-time (bandwidth-intensive) while also parsing semantic annotations (computation-intensive), creating tension between transmission speed and processing overhead. In industrial automation, robotic arms coordinating assembly operations require both sub-millisecond control signals and complex task context exchange.

\subsubsection{Real-time Development Opportunities}
Addressing real-time challenges requires architectural innovations combined with protocol redesign. 

\textbf{Integrating edge computing with asynchronous event-driven models}. For instance, multi-access edge computing (MEC) nodes can handle local synchronization and lightweight decision-making, reducing round-trip delays to under 5ms in scenarios like vehicle-to-infrastructure (V2I) coordination or industrial robotics. In smart factories, edge-deployed agents synchronize robotic arm movements with local processing, avoiding cloud latency. Asynchronous publish/subscribe mechanisms (e.g., DDS in ROS2) can further decouple dependencies, avoiding blocking behaviors during high-frequency interactions.

\textbf{Reliability-tunable communication protocols}. Building on QUIC/UDP, such protocols could offer 0-RTT connection establishment, forward error correction (FEC) for unreliable updates, and congestion-aware transmission rate control (e.g., BBR), while dynamically adapting reliability requirements. For example, emergency stop commands in industrial settings must be fully reliable, whereas temperature sensor updates can tolerate occasional loss. These protocols should support context-aware reliability selection, where critical operations automatically trigger higher reliability modes.

\subsection{High Performance}
\subsubsection{High Performance Definition and Internet Implementation}
High performance emphasizes high bandwidth utilization, massive concurrent processing, and throughput stability under extreme workloads. In systems supporting millions of concurrent users (e.g., social media platforms, online gaming), high performance ensures system stability and responsiveness during traffic spikes. For large-scale IoT deployments, high performance enables efficient data collection from millions of devices while maintaining low latency for critical operations.

The Internet achieves high performance through layered optimizations. Content Delivery Networks (CDN) distribute content globally, reducing origin server load and improving access speeds. Load balancers distribute traffic across multiple servers to prevent overload. Protocol optimizations include HTTP/2 for multiplexed requests and HTTP/3 for improved congestion control. At the network infrastructure level, techniques like traffic engineering, bandwidth reservation, and Quality of Service (QoS) mechanisms prioritize critical traffic. Cloud computing provides elastic resources that scale dynamically with demand, while edge computing reduces backhaul traffic by processing data locally.

\subsubsection{High Performance Challenges}
High performance in agent communication introduces distinct bottlenecks.

\textbf{Heterogeneous scheduling complexity}. Agents often mix RPC-style control instructions, event-driven status updates, and continuous streaming, each with divergent latency, reliability, and bandwidth requirements. For example, in smart city management, traffic control agents use RPC for intersection coordination, environmental sensors employ event-driven alerts, and surveillance cameras stream continuous video—creating conflicting demands on network resources.

\textbf{Large-scale concurrency and resource management}. At the million-agent scale (e.g., IoT clusters in smart cities), networks must allocate bandwidth fairly among urgent tasks (e.g., fire alarms) and routine tasks (e.g., environmental sensing), enforce task priorities (e.g., surgical robot commands over warehouse routing updates), and prevent congestion collapse during traffic bursts. During emergency response scenarios, thousands of agents may simultaneously request communication resources, overwhelming traditional static QoS strategies that cannot adapt effectively to such dynamic workloads.

\subsubsection{High Performance Development Opportunities}
For high performance, development opportunities include protocol, interface, and QoS standardization to ensure interoperability and guarantee resource allocation. Unified frameworks should cover diverse modes such as RPC, event-driven messaging, and streaming, with standardized formats (e.g., Protobuf serialization). Hierarchical QoS guarantees could classify flows into Platinum (e.g., urgent control $<$10ms), Gold (e.g., collaborative formation $<$50ms), and Bronze (e.g., batch analytics $<$500ms), mapping task priorities to network-level enforcement. In smart grid management, critical control commands receive Platinum QoS while routine monitoring uses Bronze resources.

\textbf{Globally optimized intelligent scheduling algorithms}. The algorithms represent another key opportunity. These algorithms can balance multiple dimensions including real-time bandwidth, latency, packet loss, and task roles (e.g., leader vs. follower). Techniques such as adaptive bitrate streaming, resource pre-reservation for critical agents, and edge caching of frequently used models (similar to CDN) can mitigate congestion. For instance, in large-scale drone swarms, leader drones receive reserved bandwidth while follower drones use adaptive rates based on network conditions.

\textbf{AI Agent Quality of Experience (QoE)}. It extends beyond human perception, emphasizing metrics such as task completion accuracy, decision timeliness, coordination efficiency, and energy/resource consumption. For example, in disaster response scenarios, QoE metrics would prioritize task completion rate (e.g., 95\% of areas surveyed within 2 hours) and resource efficiency (battery consumption per mission). By mapping QoE requirements to network-level QoS, future systems can optimize both communication infrastructure and collaborative task effectiveness, ensuring that large-scale MAS operate with both responsiveness and robustness.

\subsection{Manageability}

\subsubsection{Definition of Manageability and Internet Implementation}

Manageability refers to a system’s ability to support real-time monitoring, dynamic configuration, performance optimization, and fault diagnosis. Its foundation lies in observability (metrics, logs, traces) and controllability (APIs, policies). In distributed environments, manageability and collaboration are interdependent: manageability provides transparency and control, while efficient collaboration relies on standardized interfaces and reliable management tools. 

The Internet has long relied on layered mechanisms to ensure manageability. Monitoring is supported by protocols such as SNMP and telemetry stacks like Prometheus+Grafana; configuration is enabled through standards such as NETCONF/YANG and automated frameworks like Ansible and Terraform; and fault diagnosis is facilitated by distributed tracing systems (e.g., Jaeger) and log aggregation tools (e.g., ELK Stack). Service interoperability further benefits from HTTP/REST and gRPC, while structured formats such as JSON/XML and semantic frameworks like RDF/OWL enhance information sharing. Naming and routing mechanisms such as DNS and BGP provide global scalability and coordination. These experiences collectively demonstrate how layered design and standardization enable large-scale system manageability, and they offer valuable lessons for agent communication systems.

\subsubsection{Manageability Challenges}

Compared with the relatively structured Internet environment, agent communication presents additional manageability challenges. 

\textbf{Heterogeneous interoperability and the semantic gap}. Agents often need to exchange multimodal data—ranging from text and images to sensor streams—yet the absence of unified encapsulation standards complicates integration. For instance, autonomous driving systems must simultaneously handle high-bandwidth video feeds and structured LiDAR data, but existing protocols are not optimized for synchronized multimodal transmission. Similarly, in healthcare applications, exchanging CT images alongside diagnostic text requires differentiated QoS guarantees that are beyond the capabilities of traditional FTP/HTTP. Beyond data formats, semantic alignment is even more difficult: different agents may interpret the same concept in conflicting ways (e.g., “emergency task” may denote shutdown in industrial robots but expedited delivery in logistics robots). Syntax-based schemas such as JSON can ensure structural correctness but cannot resolve such semantic ambiguities, while ontology-based reasoning often incurs prohibitive computational overhead.

\textbf{Integration of large language models (LLMs) into agent communication}. LLM inference is computationally expensive, with typical response latencies exceeding one second, which is incompatible with real-time collaboration scenarios that demand sub-100ms responsiveness. In addition, LLM hallucinations pose serious risks in mission-critical domains, where misinterpreted instructions could lead to catastrophic consequences. Beyond inference itself, distributed context management remains difficult. Agents must maintain consistent task histories, environmental states, and role permissions across dynamic environments, but ensuring consistency, privacy, and accountability simultaneously introduces fundamental tensions that current frameworks struggle to resolve.

\subsubsection{Manageability Development Opportunities}

Opportunities for improving manageability in agent communication systems emerge along several promising directions. 

\textbf{Semantic-aware communication architectures}. Drawing inspiration from information-centric paradigms such as NDN and ICN, agents can interact based on content names rather than addresses, with edge nodes caching frequently used semantic units. Lightweight inference engines deployed at the edge (e.g., TensorFlow Lite) can perform real-time semantic annotation—such as labeling frames as “pedestrian crossing”—thus reducing central cloud workloads and improving overall observability. The standardization of multimodal data encapsulation formats, such as MPEG-V combined with JSON-LD, would further facilitate cross-agent interoperability.

\textbf{Optimizing LLM integration}. Techniques such as deploying quantized or distilled models at the edge can reduce inference latency to hundreds of milliseconds, while asynchronous pipelines can divide tasks into preprocessing, edge inference, and cloud verification to balance responsiveness with accuracy. Reliability can be strengthened through hallucination suppression strategies, including knowledge-base validation using vector databases and cross-agent consensus, in which critical decisions are only executed when confirmed by multiple heterogeneous models. Furthermore, defining explicit task-level QoE metrics—such as consistency, privacy, and traceability—and mapping them to QoS guarantees would enable systematic evaluation and management of agent collaboration.

\textbf{Incentive Alignment and Economic Models.} The long-term manageability of the agent ecosystem is inseparable from its economic model. Drawing parallels to search engines, agent discovery platforms hold immense potential for monetization, such as through sponsored capabilities (analogous to search ads) where agents pay for premium placement. This introduces a fundamental tension: how to balance monetization with the neutrality and integrity of discovery results? A key opportunity lies in designing transparent, auditable advertising models and incentive mechanisms that reward genuine value creation over mere visibility. A sustainable business model will be a cornerstone of a well-managed, trustworthy agent ecosystem.

\textbf{Configurable management frameworks and standardization}. Emerging paradigms such as policy-as-code can allow dynamic loading and enforcement of management policies at the edge, while blockchain-based auditing ensures accountability for sensitive operations. On the standardization front, semantic protocols (RDF/OWL), multimodal interaction interfaces, and LLM integration APIs (with latency SLAs and hallucination thresholds) will be critical. Together, these innovations pave the way toward configurable manageability, balancing centralized control with agent autonomy, and enabling multi-agent ecosystems to remain both flexible and governable at scale.

\section{Standardization Requirements}

As agent ecosystems evolve, standardization has become imperative for ensuring interoperability and trust. Several leading organizations have already taken initiative: the World Wide Web Consortium (W3C) has launched the AI Agent Protocol Community Group to define interoperable discovery, identity, and collaboration protocols for the Agentic Web\footnote{\url{https://www.w3.org/community/agentprotocol/}}; International Organization for Standardization (ISO) has issued standards 15067-3-30~\cite{iso-15067-3-30-2024} and 15067-3-31~\cite{iso-15067-3-31-2024}, specifying communication protocols for energy management agents; Internet Engineering Task Force (IETF) has produced multiple drafts on agent communication, such as the agent name service~\cite{narajala-ans-00}, agent authorization~\cite{rosenberg-oauth-aauth-00} and the agent identification~\cite{yl-agent-id-req-00}. These efforts underscore the urgency of establishing standards that support high performance, scalability, interoperability, and security in the agent era.
This section explores these emerging opportunities in depth.

\subsection{The Necessity and Opportunities of Standardization}

\subsubsection{Performance Standards}

In an era of agents exchanging ultra-long, multi-modal contexts and engaging in real-time, multi-agent collaboration, the demands for bandwidth, throughput, and low latency escalate dramatically. The underpinnings of communication—particularly transport protocols—must therefore evolve. Innovations such as agent-centric QUIC adaptations and quality-of-service (QoS) mechanisms can ensure efficient, secure, and context-aware routing of agent messages.

\subsubsection{Scalability Standards}

Effective discovery and interconnection of vast numbers of agents necessitate scalable naming and routing paradigms. Emerging concepts like “Agent-era DNS” or BGP-style routing for agents can provide structured, hierarchical discovery and reachability. For example, protocols such as the “Internet of Agents Protocol (IoA Protocol)” propose a layered, decentralized architecture tailored to heterogeneous agent collaboration. Likewise, “AgentDNS” offers a unified naming, resolution, and secure invocation system for LLM agents, inspired by Internet DNS.

\subsubsection{Compatibility Standards}

The diversity of agent communication protocols—ranging from ontology-based languages like FIPA-ACL and KQML to newer, generative-AI-driven protocols such as NLIP—leads to fragmentation and inefficiency. Standardizing message formats, performatives, and interaction protocols, and achieving semantic interoperability, is essential. A shared, extensible interface would allow heterogeneous agents and platforms to cooperate seamlessly.

\subsubsection{Security Standards}

Agent collaboration introduces security challenges absent in traditional networks—such as cascading authorization, dynamic delegation, and privacy-sensitive context sharing. Developing formal security protocols and standards for identity authentication, capability attestation, delegation constraints, and confidentiality guarantees is indispensable for safeguarding large-scale agent ecosystems.

\subsection{Layers for Standardization}

\subsubsection{Lower-Layer (Transport Layer) Standardization}

The transport layer encompasses foundational protocols responsible for reliable, orderly, and efficient data exchange between endpoints. Core protocols such as TCP and UDP have served this role for decades, providing error correction, flow control, and multiplexing capabilities. In recent years, QUIC has emerged as a modern addition to this layer: a UDP-based, multiplexed protocol engineered to reduce latency, avoid head-of-line blocking, and resist protocol ossification through encrypted metadata and flexible extension mechanisms.

Standardizing at this layer yields notable advantages. A unified, efficient transport substrate ensures that diverse agent systems benefit from consistent performance characteristics and baseline security guarantees. QUIC’s design, for instance, demonstrates how transport-layer evolution can deliver improved latency and extensibility without requiring kernel-level modifications. Moreover, broad adoption of such standards can facilitate interoperable and high-throughput agent-to-agent exchanges in distributed settings.

Nonetheless, there are inherent challenges. Creating consensus around new transport standards is a slow, multi-stakeholder endeavor and may lag behind rapid shifts in agent communication needs. Once entrenched, these standards often exhibit “ossification,” making future extension or adaptation difficult. Historical trends—such as the limited adoption of the OSI model, versus the pragmatic dominance of TCP/IP—highlight the danger that rigid bottom-up standards may stifle innovation rather than support it.

\subsubsection{Upper-Layer (Application Layer) Standardization}

Application-layer standardization concerns the syntax, semantics, and protocols that enable agents to understand, coordinate, and invoke each other’s capabilities. Emerging standards like the Model Context Protocol (MCP), Agent Communication Protocol (ACP), Agent-to-Agent Protocol (A2A), and Agent Network Protocol (ANP) address cross-agent interoperability by defining structured messaging formats, discovery mechanisms, and peer-to-peer orchestration frameworks.

Standardizing at the application layer provides agility and responsiveness to domain-specific use cases. For example, ACP enables REST-native, asynchronous, multi-modal communication, while A2A facilitates peer-to-peer task delegation using capability-based cards—allowing rapid evolution of agent workflows tailored for collaborative pipelines and enterprise-scale deployment. This flexibility empowers developers to quickly adapt communication protocols to new agent collaboration forms without waiting for slow consensus cycles.

However, such modular autonomy introduces risks of fragmentation. Without coordination across industries and platforms, application-level standards may proliferate into isolated “protocol islands,” undermining interoperability. Agents built on different standards may struggle to discover or collaborate with one another, negating many benefits of standardization. As the landscape evolves, balancing standard flexibility with ecosystem coherence becomes a central governance challenge.

\subsection{Competition for Agent Communication Entry Points}

Beyond technical standardization, a key strategic issue is the competition among different players for control of the primary entry points of agent communication, reminiscent of past struggles over browsers, search engines, and super-apps. Hardware and OS vendors seek dominance by embedding default agents, while application platforms leverage user bases and network effects, though both face constraints from ecosystem dependencies. Cloud providers and model companies contribute advanced AI capabilities but lack direct consumer interfaces, whereas telecom operators rely on network and identity control yet struggle with weak user engagement. These competitions map onto different layers of standardization: hardware vendors and operators favor \textit{lower-layer standards} (e.g., transport protocols, identity infrastructures) to retain performance and reliability control; application platforms and cloud providers push for \textit{upper-layer standards} (e.g., compatibility, semantic interoperability) to differentiate services; model providers emphasize \textit{performance and scalability standards}; and various stakeholders promote \textit{security standards} to strengthen trust and governance. 
The outcome of this race for entry points will shape the direction of standardization: dominant players may push proprietary protocols that fragment ecosystems, while open standards promise interoperability but face slower adoption.

\section{Conclusion}

In this paper, we reviewed the evolution of agent communication technologies and distilled five key elements from Internet architecture: scalability, security, real-time performance, high performance, and manageability. Using these elements, we analyzed the main challenges and opportunities of agent communication and outlined potential standardization pathways. Nonetheless, this study has limitations: it does not fully cover fast-evolving technologies, practical deployment cases remain limited, and the impact of ongoing standardization efforts has yet to be validated. Future work should broaden the scope of technology review, integrate empirical case studies, and evaluate the outcomes of emerging standardization efforts.



\balance
\bibliography{references}
\bibliographystyle{IEEEtran}

\end{document}